\documentclass[aps,preprint]{revtex4-2}

\usepackage[latin1]{inputenc}
\usepackage[T1]{fontenc}
\usepackage{amsmath}
\usepackage{amsthm}
\usepackage{amssymb}
\usepackage{hyperref}
\usepackage{dsfont}
\usepackage{graphicx}
\usepackage{tikz-cd}
\usepackage{blindtext}

\hypersetup{colorlinks=true, linkcolor=blue, citecolor=blue}

\begin{document}
\title{Gap modes in Arnold tongues and their topological origins}

\begin{abstract}
Gap modes in a modified Mathieu equation, perturbed by a Dirac delta potential, are investigated. It is proved that the modified Mathieu equation admits stable isolated gap modes with topological origins in the unstable regions of the Mathieu equation, which are known as Arnold tongues. The modes may be identified as localized electron wavefunctions in a 1D chain or as toroidal Alfv\'en eigenmodes. A generalization of this argument shows that gap modes can be induced in regimes of instability by localized potential perturbations for a large class of periodic Hamiltonians.    
\end{abstract}

\author{Andrew Brown}
\email{aobrown@princeton.edu}
\author{Hong Qin}
\email{hongqin@princeton.edu}
\affiliation{Department of Astrophysical Sciences, Princeton University, Princeton, New Jersey 08544}
\affiliation{Plasma Physics Laboratory, Princeton University, Princeton, NJ 08543, U.S.A}

\maketitle

Recent work has shown that the theory of topological insulators - as first explored in condensed matter physics \citep{hasan_colloquium_2010} - is relevant to continuum physics as well. Previously understood equatorial modes have been shown to have a topological origin \citep{delplace_topological_2017}, and novel modes in plasmas have also been predicted using the same tools \citep{parker_nontrivial_2020,qin_topological_2023}. Here we will elucidate the topological origins of a well-known plasma mode, the TAE \citep{cheng_high-n_1985, cheng_lown_1986}, by introducing a toy model for a perturbed Mathieu equation, and showing that it exhibits a localized mode whose existence is insensitive to the details of the perturbation.

The Mathieu equation
\begin{equation}
    \frac{d^2 x}{dt^2}+(\delta +\epsilon \cos t)x = 0
    \label{mathieu}
\end{equation}
arises naturally in the study of acoustics \citep{Mathieu1868}, condensed matter physics \citep{carver_mathieus_1971}, and plasma physics \citep{cheng_high-n_1985}, among other fields. Since it is the equation of motion for a modified harmonic oscillator with a periodic fundamental frequency, the Mathieu equation is prototypical in the study of periodic Hamiltonian systems. When $\epsilon$ is nonzero, the stability chart of the Mathieu equation shows regions of instability, sometimes called Arnold tongues, separating the continuous spectrum \citep{kovacic_mathieus_2018}. 

In this work we are motivated by a modified form of the Mathieu equation arising from the ballooning representation of toroidal Alfv\'en eigenmodes (TAE) in the limit of high toroidal mode number \citep{cheng_high-n_1985}. These global eigenmodes are observed to lie in the continuum gaps of the magnetohydrodynamic spectra of toroidal systems. Because such gap modes may be excited by energetic particles (e.g., fusion products) and do not undergo continuum damping, they are of significant interest in the study of magnetic confinement fusion \citep{chen_physics_2016}.

The authors of \citep{cheng_high-n_1985} examined the large-aspect-ratio (small-$\epsilon$) asymptotics of the gap mode arising from a certain choice of perturbation to the Mathieu equation. In this work we generalize their efforts, showing that a large class of temporally localized perturbations to the Mathieu equation will generate a gap mode. Moreover, our argument shows that a gap mode may exist in any continuum gap in any periodic Hamiltonian system for suitable choice of perturbation. 

The paper is organized as follows: We first motivate the choice of a Dirac delta perturbation and prove the existence of a gap mode. We then explore the small-$epsilon$ asymptotics of the gap mode and compare them to numerical results. Finally, we argue from numerical evidence that the qualitative behavior of the gap mode is relatively insensitive to the width or functional form of the perturbation.

We conclude with some comments on the implications for this work for the study of TAE and other Alfv\'en eigenmodes, in addition to some speculation about generalizations of our work to perturbations of finite width.

{\bf Analytical Description of Gap Mode.} We demonstrate that a delta-function perturbation to the Mathieu equation generates a gap mode in each of the regions of instability.

In the seminal work of Cheng and Chance on  TAE in the high-toroidal-mode-number limit in which the ballooning formalism may be applied, the governing equation for TAE is a modified form of Eq.~\eqref{mathieu}, with a non-periodic term added to the potential:
\begin{equation}
    \frac{d^2 x}{dt^2}+(\delta +\epsilon \cos t + F(t))x = 0.
    \label{mod-mathieu}
\end{equation}
In their work, the function $F(t)$ takes the form
\begin{equation}
    F(t) = -\frac{s^2}{(1+s^2 t^2)^2},
\end{equation}
where $s$ is related to the magnetic shear of the equilibrium profile of a tokamak. Recalling the representation of the Dirac $\delta$ function as the limit
\begin{equation}
    \delta (t) = \lim_{\eta \to 0}\frac{1}{\pi}\frac{\eta}{\eta^2 + t^2},
\end{equation}
and in pursuit of a tractable equation, we take $F(t)=-\lambda \delta(t)$. It may be insightful to compare the resulting differential equation to some equivalent quantum mechanical system - a harmonic oscillator with a periodic variation that is ``kicked," briefly, at $t=0$, or an electron in a 1D lattice with some localized impurity. It is well-known that, for some values of $\delta$ and $\epsilon$, Eq.~\eqref{mathieu} is unstable, so solutions grow without bound at $t\to \pm \infty$. We say that there is a ``gap" in the spectrum of the Mathieu equation. However, using a familiar technique from quantum mechanics to study a Dirac delta potential, we can tune $\lambda$ to construct a special mode that decays at $t \to \pm \infty$ even though $\delta$ is in the continuum gap.

\begin{figure}
    \centering
    \includegraphics[width=0.42\textwidth]{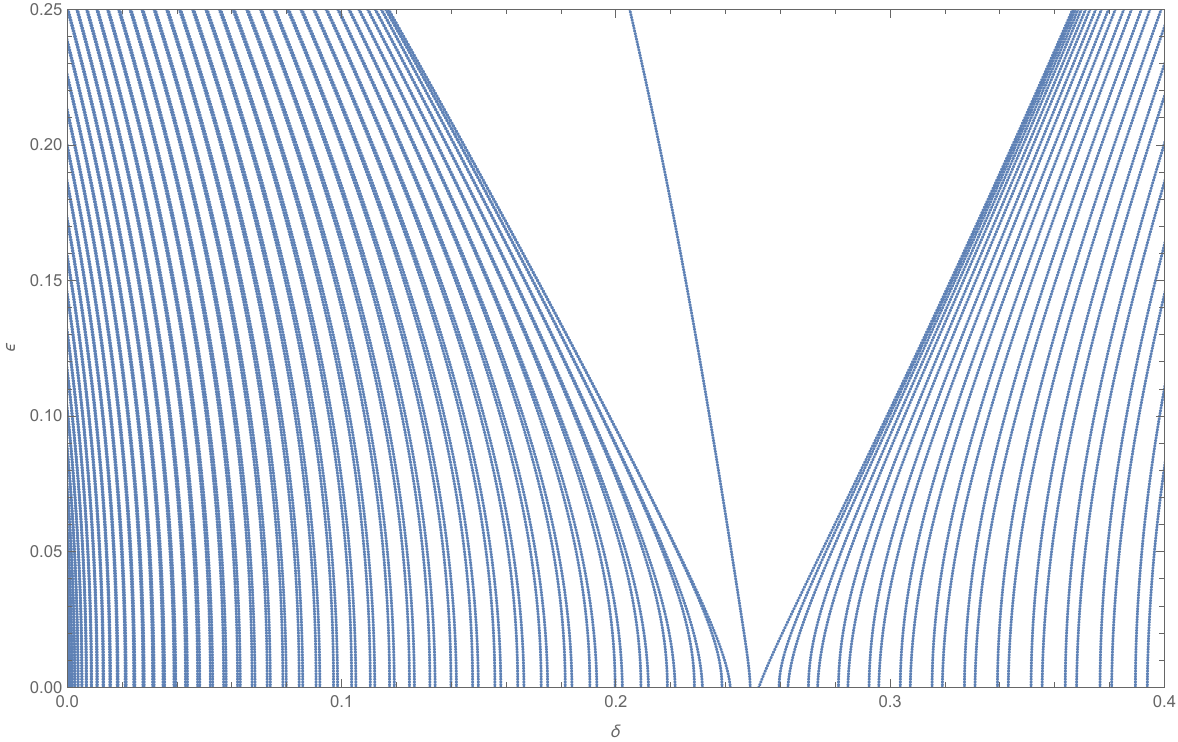}
    \caption{A gap mode in the Arnold tongue around $\delta=0.25$, $\lambda = 0.7$.}
    \label{025gap}
\end{figure}

\begin{figure}
    \centering
    \includegraphics[width=0.47\textwidth]{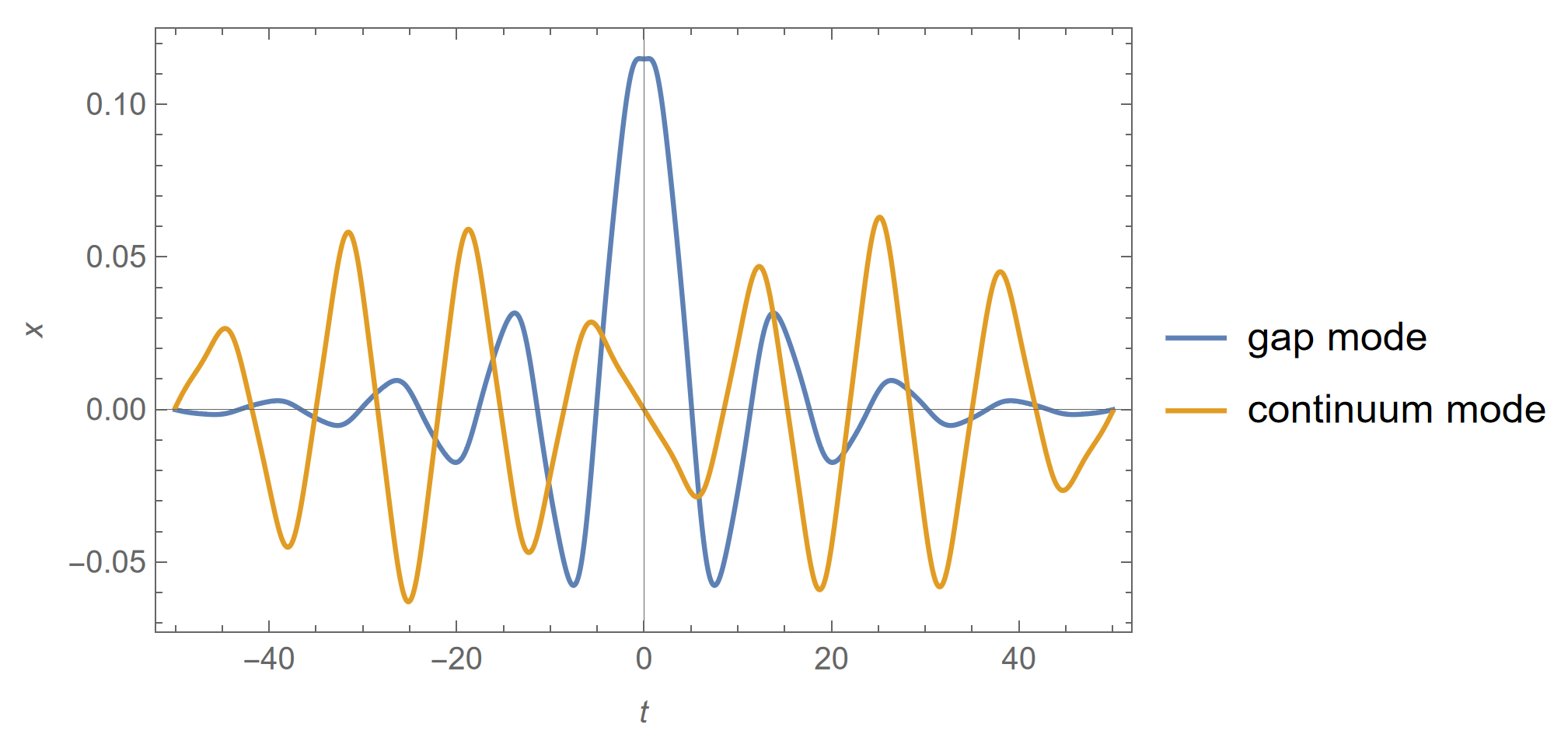}
    \caption{A comparison of a decaying gap mode to a quasiperiodic mode on the edge of the continuum gap. }
    \label{gap-vs-nongap}
\end{figure}

\begin{figure}
    \centering
    \includegraphics[width=0.42\textwidth]{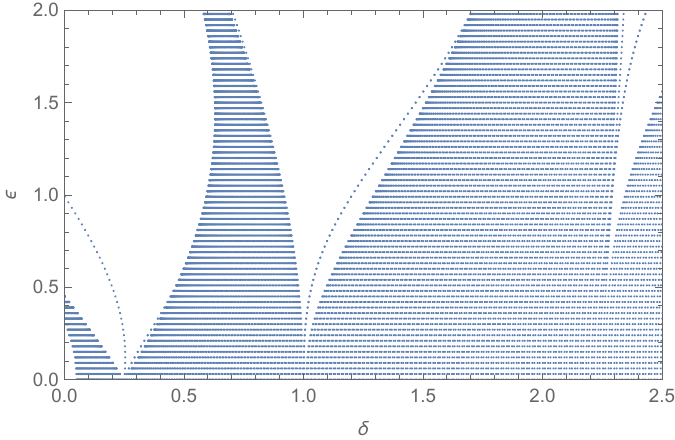}
    \caption{Gap modes in the Arnold tongues around $\delta=0.25\text{, }1.0\text{, and }2.25$, $\lambda = 1.0$.}
    \label{multigap}
\end{figure}

Let $\epsilon$ and $\delta$ be fixed, such that $\delta$ is in the continuum gap of  Eq.~\eqref{mathieu}. Since the Mathieu equation is a second-order linear ODE, there are two linearly independent solutions. One is an unstable mode in (showing unbounded growth as $t\to +\infty$) with some complex frequency $\nu$. Since the Mathieu equation is arises from a Hamiltonian, we know that the second mode has characteristic frequency $\nu^*$, and this mode decays at $t \to +\infty$. On the other hand, these two modes must show opposite behavior at $t \to -\infty$. Denote by $m_{\pm}(t)$ the solution that decays at $t\to \pm\infty$, respectively, so a general solution to Eq.~\eqref{mod-mathieu} is 
\begin{equation}
    x_{\pm}(t) = \alpha_{\pm} m_{-}(t) + \beta_{\pm} m_{+}(t)
\end{equation}
where $\pm$ corresponds to $t>0$ and $t<0$, respectively. For finiteness at $t\to \pm \infty$, $\beta_{-}=\alpha_{+}=0$. For continuity at $t=0$,
$\alpha_{-}m_{-}(0) = \beta_{+}m_{+}(0)$. And by integrating over an infinitesimal interval containing the origin,
\begin{equation}
    \beta_{+}m_{+}'(0) - \alpha_{-}m_{-}'(0) = \lambda \beta_{+}m_{+}(0).
\end{equation}
Substituting the continuity equation and dividing through by $\alpha_{-}m_{-}(0)$, we find a requirement on $\lambda$ for such a bounded gap mode to exist:
\begin{equation}
    \lambda = \frac{m_{+}'(0)}{m_{+}(0)}-\frac{m_{-}'(0)}{m_{-}(0)}.
    \label{lam-req}
\end{equation}
Even though we do not have an explicit form for $m_{\pm}$, this expression does tell us that, given a pair $(\epsilon, \delta)$, there is precisely one value of $\lambda$ that yields a gap mode. This is very similar to the familiar case of the delta function potential for the Schrodinger equation, in which a single bound state exists. We can also make several qualitative statements about the gap modes before moving on to asymptotics in the next section. First, we can numerically check for the existence of a mode in the continuum gap; the presence of the predicted mode is unambiguous in Fig.~\ref{025gap}. Since the $m_{\pm}$ are decaying at $\pm \infty$ by definition, the gap mode will be localized near zero (or wherever the delta function kick is placed), in contrast to the quasiperiodic states on the edge of the continuum gap. This behavior is clearly borne out by Fig.~\ref{gap-vs-nongap}. And the preceding arguments apply in any region of instability for the Mathieu equation, so we should be able to see gap modes in all of the Arnold tongues of the Mathieu equation. These gaps open from $\delta = n^2/4$ for $n=1,2,3,...$, and Fig.~\ref{multigap} shows a gap mode in the first three of these gaps. 

Incidentally, it is simple to show that Eq.~\eqref{lam-req} disallows gap modes when $\lambda=0$. Consider the unmodified Mathieu equation as an IVP, starting at $t=0$. Then 
\begin{equation}
    x(t) = \alpha m_{+}(t) + \beta m_{-}(t)
\end{equation}   
and we take $x(0)$ and $x'(0)$ to be known constants, so the constants $\alpha$ and $\beta$ are determined from the matrix equation
\begin{equation}
    \begin{pmatrix}
    m_{+}(0) & m_{-}(0)\\
    m_{+}'(0) & m_{-}'(0)
    \end{pmatrix}
    \begin{pmatrix}
    \alpha \\ \beta
    \end{pmatrix} =
    \begin{pmatrix}
    x(0) \\ x'(0)
    \end{pmatrix}.
\end{equation}
Taking the determinant and comparing to Eq.~\eqref{lam-req}, we see
\begin{equation}
\begin{aligned}
    m_{+}(0) m_{-}'(0) -& m_{-}(0)m_{+}'(0)\\ = -&\lambda m_{+}(0) m_{-}(0),
\end{aligned}
\end{equation}
so the determinant vanishes and the IVP is degenerate when $\lambda=0$. Asymptotic analysis shows that there is also no gap mode for $\lambda < 0$ in the limit of asymptotically small $\epsilon$.

{\bf Asymptotics.}
We now study the asymptotic behavior of the gap mode spectrum at small $\epsilon$ in the gap near $\delta \sim 0.25$. 

First, we find the slow decay of the gap mode using the method of multiple scales. Following Rand and Kovacic, we define fast and slow variables $t$ and $\eta = \epsilon t$. Then the kicked Mathieu equation becomes
\begin{equation}
\begin{aligned}
    \frac{\partial^2 x}{\partial t^2} +& 2 \epsilon \frac{\partial^2 x}{\partial t \partial \eta} + \epsilon^2 \frac{\partial^2 x}{\partial \eta^2} \\
    &+ (\delta + \epsilon \cos t)x = \lambda \delta(t) x.
\end{aligned}
\end{equation}
We also expand the solution $x$ and the eigenvalue $\delta$ in asymptotic series
\begin{equation}
    x\sim x_0 + \epsilon x_1 + \mathcal{O}(\epsilon^2)
\end{equation}
and
\begin{equation}
    \delta \sim \frac{1}{4} + \epsilon \delta_1 + \mathcal{O}(\epsilon^2).
\end{equation}
Then the lowest order solution is a sum of fast oscillations that grow or decay on the slow timescale,
\begin{equation}
\begin{aligned}
    x_0& = A(\eta)\cos(\sqrt{\delta}t) + B(\eta)\sin(\sqrt{\delta}t)\\
    &\simeq A(\eta)\cos\left(\frac{t}{2}\right) 
    + B(\eta)\sin\left(\frac{t}{2}\right).
\end{aligned}
\end{equation}
 The first-order equation is
\begin{equation}
\begin{aligned}
    \frac{\partial^2 x_1}{\partial t^2}& + \frac{1}{4}x_1 + 2\frac{\partial^2 x_0}{\partial t \partial \eta} \\+ &\left(\cos t  + \delta_1\right)x_0 - \lambda \delta(t) x_1 = 0.
\end{aligned}
\end{equation}
 We find resonant terms (which are proportional to $\sin (t/2)$ or $\cos (t/2)$. Setting their coefficients to zero gives the slow amplitudes $A$ and $B$, which depend on $\delta_1$ as
\begin{equation}
\begin{aligned}
    A(\eta) &= A_1 \exp\left(\sqrt{\frac{1}{4}-\delta_1^2}\eta\right)\\
    +A_2 &\exp\left(-\sqrt{\frac{1}{4}-\delta_1^2}\eta\right),
\end{aligned}
\end{equation}
for some constants $A_{1,2}$, and similarly for $B(\eta)$.
Thus the lowest order behavior of the gap mode follows
\begin{equation}
    x_{0,\pm}(t) \sim \exp\left(\mp \sqrt{\frac{1}{4}-\delta_1^2}\epsilon t\right).
    \label{decay}
\end{equation}
We could proceed in the usual way to the next order in $\epsilon$ via the method of multiple scales, eventually finding a non-resonance condition that yields a formula for $\delta_1$. It turns out that it suffices to preemptively extract the slow decay described by Eq.~\eqref{decay} and then $\delta_1$ is determined at lowest order in $\epsilon$ in the equation for $y(t)$, defined by
\begin{equation}
    x(t) = \exp\left(\mp \sqrt{\frac{1}{4}-\delta_1^2}\epsilon t\right) y(t).
\end{equation}
We asymptotically expand $y\sim y_0 + \epsilon y_1$, yielding two equations (for $t<0$ and $t>0$),
\begin{equation}
\begin{aligned}
    \frac{d^2 y}{dt^2} &\mp 2 \epsilon \sqrt{\frac{1}{4}-\delta_1^2}\frac{dy}{dt} + \epsilon^2 \left(\frac{1}{4}-\delta_1^2\right)y \\+ &\left(\frac{1}{4} + \epsilon \delta_1 +\epsilon \cos(t)\right) y = 0
\end{aligned}
\end{equation}
and the solutions $y_{\pm}$ must obey the transition condition
\begin{equation}
    \lambda = \frac{y_{+}'(0)}{y_{+}(0)}-\frac{y_{-}'(0)}{y_{-}(0)}.
    \label{jump-cond}
\end{equation}
Then at lowest order
\begin{equation}
    y_0 = A_{\pm}\cos (t/2)+ B_{\pm}\sin (t/2)
\end{equation}
and the next order equation
\begin{equation}
\begin{aligned}
    \frac{d^2 y_1}{dt^2}& + \frac{1}{4}y_1 =\\ \mp &\sqrt{\frac{1}{4}-\delta_1^2}(A_{\pm}\sin(t/2)
    -B_{\pm}\cos(t/2))\\-&(\delta_1 + \cos t)(A_{\pm}\cos (t/2)+ B_{\pm}\sin (t/2)).
\end{aligned}
\end{equation}
Eliminating terms exhibiting secular growth gives the conditions
\begin{equation}
    \frac{B_{\pm}}{A_{\pm}} = \mp\frac{\sqrt{1-4\delta_1^2}}{2\delta_1 -1}.
\end{equation}
Inserting into the jump condition from Eq.~\eqref{jump-cond} gives
\begin{equation}
    \lambda = \frac{2 \sqrt{1-4 \delta_1^2}}{1-2\delta_1}.
\end{equation}
Since the RHS is always positive (for all $\delta_1$ inside the gap), there is only a gap mode if $\lambda$ is positive as well. 

\begin{figure}
    \centering
    \includegraphics[width=0.42\textwidth]{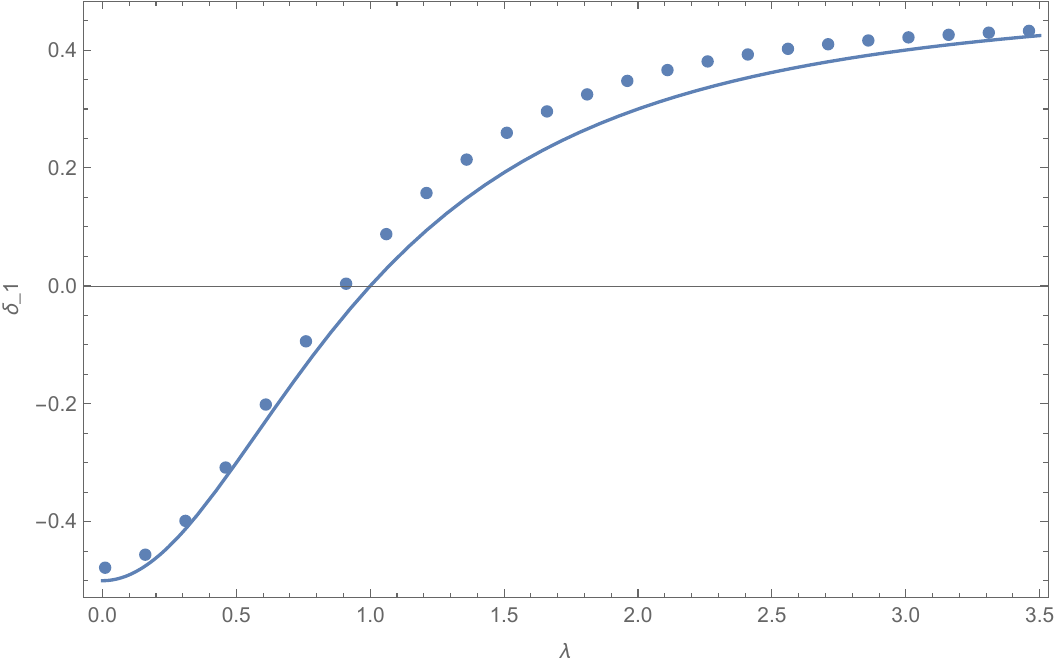}
    \caption{Points are numerically calculated values of $\delta_1$. The solid curve is the analytical form from Eq.~\eqref{d1-anal}.}
    \label{d1-vs-lambda}
\end{figure}

And, solving for $\delta_1$,
\begin{equation}
    \delta_1 = \frac{\pm \lambda^2 - 1}{2 (1+\lambda^2)},
    \label{d1-anal}
\end{equation}
where the plus sign gives the gap mode.
Note that at $\lambda=0$, $\delta_1$ tends to $-1/2$, which is one boundary of the gap. At large $\lambda$, $\delta_1$ tends to $1/2$, the other boundary of the gap. This asymptotic behavior is in good agreement with numerical estimates, as shown in Fig.~\ref{d1-vs-lambda}.

Though the gap mode is even by construction, as $\lambda$ tends to $\infty$ as $\delta_1$ approaches the upper boundary of the gap, the gap mode on the right half-line tends toward the \textit{odd} continuum mode. The singularity in $\lambda$ as the gap mode transits between modes of different parity is homotopically retained from the limit of asymptotically small $\epsilon$ up to finite values of $\epsilon$, so the gap mode persists.

\begin{figure}
    \centering
    \includegraphics[width=0.42\textwidth]{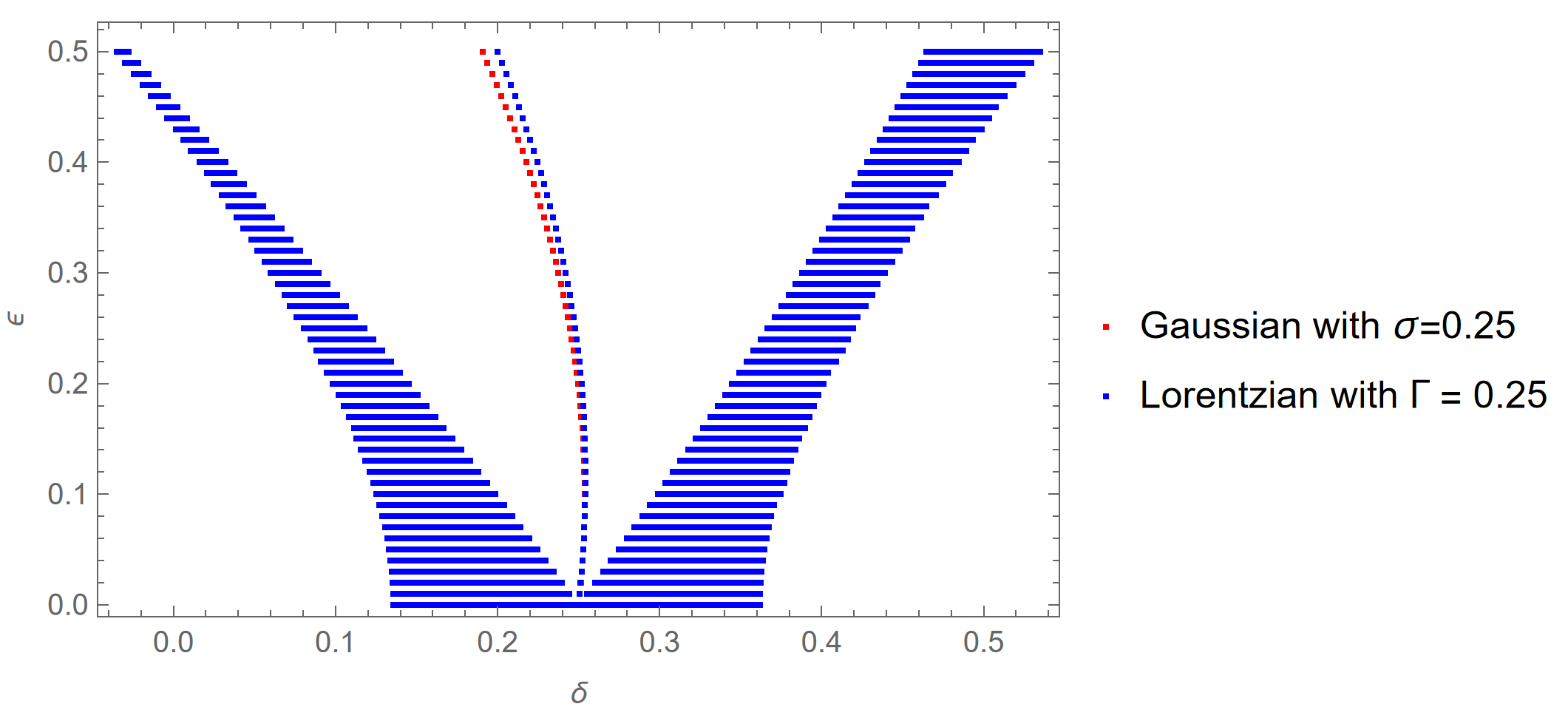}
    \caption{Spectra for Gaussian (blue) and Lorentzian (orange) perturbations with the same width $0.25$, $\lambda = 1.0$.}
    \label{gaussian-vs-lorentzian}
\end{figure}

{\bf Modifications to the Perturbation.}
We verify numerically that the gap modes described above, arising from a Dirac delta perturbation, persist for broader approximations to a delta distribution, including Gaussian and Lorentzian kicks. As anticipated, when the perturbation width is small (as in Fig.~\ref{gaussian-vs-lorentzian}) the spectrum is insensitive to the functional form of the perturbation.


\begin{figure}
    \centering
    \includegraphics[width=0.42\textwidth]{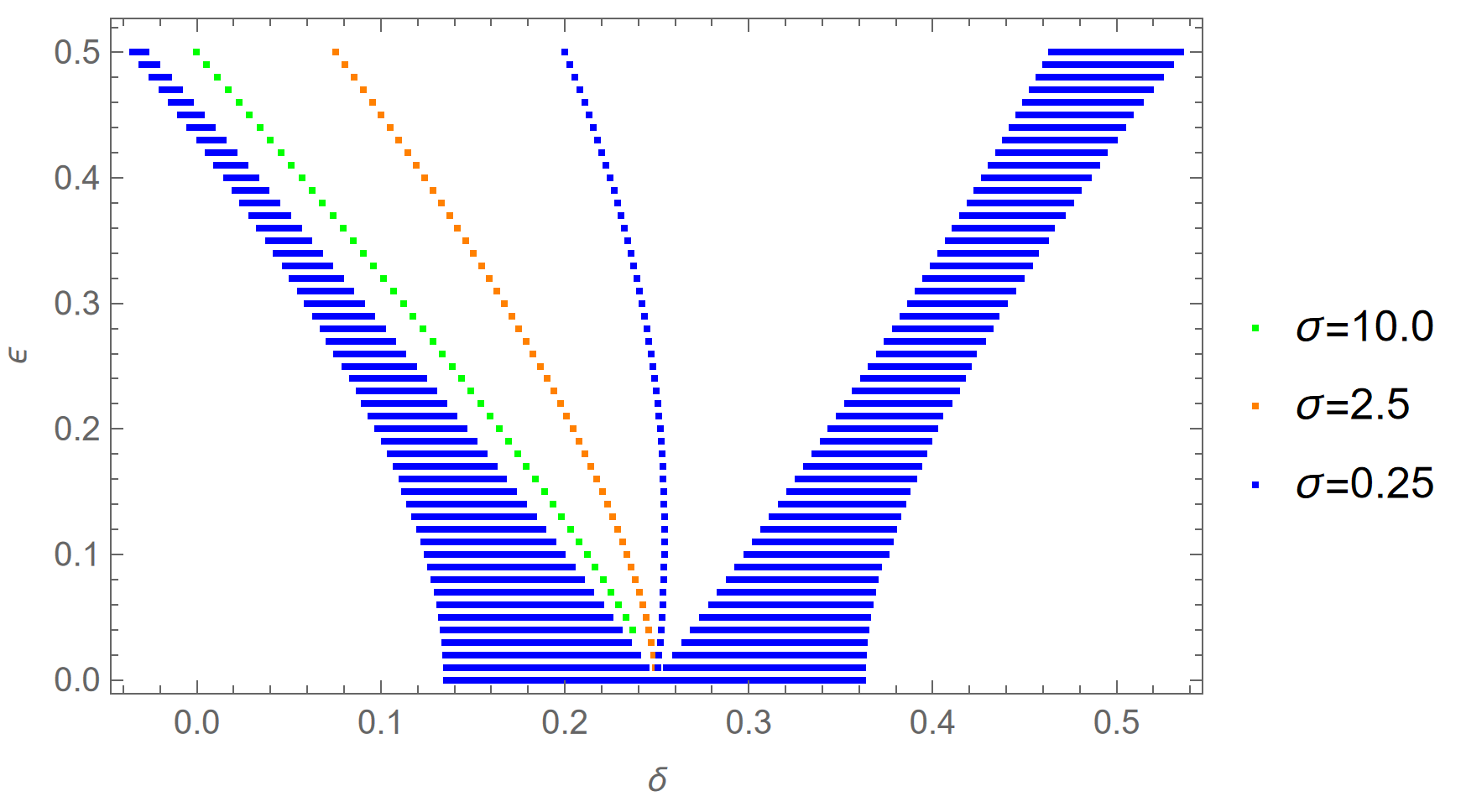}
    \caption{Numerical spectra with varying width of Gaussian perturbation, with $\lambda = 1.0$.}
    \label{kick-width}
\end{figure}

{\bf Conclusion.} We have presented a novel modification of the Mathieu equation that admits an isolated mode at every gap in the continuum spectrum. This mode is now well understood at asymptotically small values of $\epsilon$, and, by a parity argument, is topologically protected at finite values of $\epsilon$.

This mode may be interpreted physically as a spatially localized particle, induced by an impurity in a 1D lattice, or as an idealization of the TAE. It remains to be seen whether a similar approach can be used to find gap modes in higher-dimensional lattices or non-axisymmetric magnetic configurations.

\begin{acknowledgments}
This work is supported by U.S. Department of Energy (DE-AC02-09CH11466).
\end{acknowledgments}


\bibliographystyle{apsrev4-2}
\bibliography{topolit}

\end{document}